\documentclass[%
reprint,
superscriptaddress,
showpacs,preprintnumbers,
 amsmath,amssymb,
 aps,
 longbibliography,
pra,
]{revtex4-2}
\usepackage[utf8]{inputenc}
\usepackage[thinspace]{SIunits}
\usepackage{xspace}
\usepackage[final]{graphicx}
\usepackage{dcolumn}% Align table columns on decimal point
\usepackage{bm}% bold math
\usepackage{color}
\usepackage{soul}
\usepackage{multirow}
\usepackage{hyperref}
 \usepackage{amssymb}
 \usepackage{soul}

\def\adda#1{}

\def\maa#1{}

\hypersetup{
    unicode=false,          % non-Latin characters in Acrobat’s bookmarks
    pdftoolbar=true,        % show Acrobat’s toolbar?
    pdfmenubar=true,        % show Acrobat’s menu?
    pdffitwindow=false,     % window fit to page when opened
    pdfstartview={FitH},    % fits the width of the page to the window
    pdftitle={Raman mode symmetry and competing phases in Mn3Si2Te6},    % title
    pdfnewwindow=true,      % links in new window
    colorlinks=true,       % false: boxed links; true: colored links
    linkcolor=blue,          % color of internal links (change box color with linkbordercolor)
    citecolor=blue,        % color of links to bibliography
    filecolor=blue,      % color of file links
    urlcolor=blue,       % color of external links
    plainpages=false,        % hat Einfluss auf die Seitennummerierung
}

\makeatletter
\DeclareRobustCommand{\textsupsub}[2]{{%
  \m@th\ensuremath{%
    ^{\mbox{\fontsize\sf@size\z@#1}}%
    _{\mbox{\fontsize\sf@size\z@#2}}%
  }%
}}
\makeatother

\newcommand{\Tst}{\ensuremath{T^\mathrm{*}}\xspace}

\newcommand{\CST}{\mbox{CrSiTe$_3$}\xspace}

\newcommand{\MnSiTe}{\mbox{Mn$_3$Si$_2$Te$_6$}\xspace}

\newcommand{\Alg}{\texorpdfstring{\ensuremath{A_{1g}}\xspace}{A1g}}

\newcommand{\Asu}{\texorpdfstring{\ensuremath{A_{2u}}\xspace}{A2u}}

\newcommand{\Eg}{\texorpdfstring{\ensuremath{E_{g}}\xspace}{Eg}}

\newcommand{\Eu}{\texorpdfstring{\ensuremath{E_{u}}\xspace}{Eu}}

\usepackage{array}
\newcolumntype{L}[1]{>{\raggedright\let\newline\\\arraybackslash\hspace{0pt}}m{#1}}

\newcommand{\wn}{\ensuremath{\rm cm^{-1}}\xspace}
\begin{document}

%%%%%%%%%%%%%%%%%%%%%%%%%%%%%%%%%%%%
%\begin{CJK*}{GBK}{}%%%%%%%%%%%%%%%%%
%%%%%%%%%%%%%%%%%%%%%%%%%%%%%%%%%%%%

\title{Spin-phonon interaction and short range order in $\MnSiTe$}
\date{\today}

\author{S. Djurdji\'{c} Mijin}
\affiliation{Institute of Physics Belgrade, University of Belgrade, Pregrevica 118, 11080 Belgrade, Serbia}
\author{A. \v{S}olaji\'{c}}
\affiliation{Institute of Physics Belgrade, University of Belgrade, Pregrevica 118, 11080 Belgrade, Serbia}
\author{J. Pe\v{s}i\'{c}}
\affiliation{Institute of Physics Belgrade, University of Belgrade, Pregrevica 118, 11080 Belgrade, Serbia}
\author{Y.~Liu}
\altaffiliation{Present address: Los Alamos National Laboratory, Los Alamos, New Mexico 87545, USA}
\affiliation{Condensed Matter Physics and Materials Science Department, Brookhaven National Laboratory, Upton, NY 11973-5000, USA}

\author{C.~Petrovic}
\affiliation{Condensed Matter Physics and Materials Science Department, Brookhaven National Laboratory, Upton, NY 11973-5000, USA}
\author{M. Bockstedte}
\affiliation{Institute for Theoretical Physics, Johannes Kepler University Linz, Altenbergerstr. 69, 4040 Linz, Austria}
\author{A.~Bonanni}
\affiliation{Institute of Semiconductor and Solid-State Physics, Johannes Kepler University Linz, Altenbergerstr. 69, 4040 Linz, Austria}
\author{Z. V.~Popovi\'{c}}
\affiliation{Institute of Physics Belgrade, University of Belgrade, Pregrevica 118, 11080 Belgrade, Serbia}
\affiliation{Serbian Academy of Sciences and Arts, Knez Mihailova 35, 11000 Belgrade, Serbia}
\author{N.~Lazarevi\'{c}}
\affiliation{Institute of Physics Belgrade, University of Belgrade, Pregrevica 118, 11080 Belgrade, Serbia}
%

%%%%%%%%%%%%%%%%%%%%%%%%%%%%%%%%%%%%%%%%%%%%%%%%%%%%%%%%%%%%%%%%%%%%%%%%%%%%
\begin{abstract}
The vibrational properties of ferrimagnetic \MnSiTe single crystals are investigated using Raman spectroscopy and density functional theory calculations. Eighteen Raman-active modes are identified, fourteen of which are assigned according to with the trigonal symmetry. Four additional peaks, obeying the $\Alg$ selection rules, are attributed to the overtones. The unconventional temperature evolution of the $\Alg^5$ mode self-energy suggests a competition between different short-range magnetic correlations that significantly impact the spin-phonon interaction in \MnSiTe. The research provides a comprehensive insight to the lattice properties, studies their temperature dependence and shows the arguments for existence of competing short-range magnetic phases in \MnSiTe. 
 \end{abstract}
%%%%%%%%%%%%%%%%%%%%%%%%%%%%%%%%%%%%%%%%%%%%%%%%%%%%%%%%%%%%%%%%%%%%%%%%%%%%%%%%%%%%%%%%%%%%%%%
\pacs{%
%78.30.-j, %Infrared and Raman spectra
%74.72.-h, %cuprate superconductors
%74.70.Xa, %pnictides and chalcogenides
%75.10.Jm, %quantized spin models including frustration
%74.20.Mn, %nonconventional mechanisms
%74.25.nd %Raman and optical spectroscopy (of superconductors)
}
\maketitle

%%%%%%%%%%%%%%%%%%%%%%%%%
%\end{CJK*}%%%%%%%%%%%%%%
%%%%%%%%%%%%%%%%%%%%%%%%%

%%%%%%%%%%%%%%%%%%%%%%%%%%%%%%%%%%%%%%%%%%%%%%%%%%%%%%%%%%%%%%%%%%%%%%%%%%%%%%%%%%%%%%%%%%%%

\section{Introduction}

Layered magnetic van der Waals materials have lately received widespread attention due to their potential application in spintronics, magneto-electronics, data storage and biomedicine \cite{condmat5020042, doi.org/10.1038/nnano.2012.193, Han_2014, Zhang_2019, doi:10.1021/ac5039863, https://doi.org/10.1118/1.2806958, L1}. Recent experimental confirmation of a long-range magnetism persisting down to a monolayer in CrI$_3$ \cite{B.Huang2017_N546_270-273} further affirmed a these materials as  platform for magneto-optoelectronic devices \cite{Jiang_2018}, and as candidates for studying low-dimensional magnetism \cite{10.1016/j.mattod.2019.03.015}. 

$\MnSiTe$ single crystals were first synthesized in 1985 \cite{VINCENT1986349}. However, few studies were carried out on this compound since. It was only recently that the attention has shifted to them, mainly through comparisons with quasi-two-dimensional materials, specifically $\CST$. The vast majority of recent studies were focused on explaining the magnetism in $\MnSiTe$ and determining its crystal structure. It was revealed that $\MnSiTe$ crystallizes in a trigonal structure described by the $P\bar{3}1c$ (No. 163) space group \cite{VINCENT1986349, PhysRevB.98.064423}. According to various magnetization studies, $\MnSiTe$ is an insulating ferrimagnetic with Currie temperature $T_c$ between 74\;K --78\;K \cite{RIMET19817, PhysRevB.98.064423, doi:10.1063/5.0002168, PhysRevB.95.174440}. First principle calculations suggested a competition between ferrimagnetic ground state and three additional magnetic configurations, originating from antiferromagnetic exchange for the three nearest Mn-Mn pairs \cite{PhysRevB.95.174440}. Additionally, both magnetization and diffuse neutron scattering experiments point at the existence of strong spin correlations well above $T_c$, which may be associated with short-range order or to the preserved correlated excitations in the paramagnetic region \cite{PhysRevB.98.064423, PhysRevB.95.174440}. 

Here we present an experimental and theoretical Raman scattering study of $\MnSiTe$ single crystals, with the focus on phonon properties in the temperature range from 80\;K to 320\;K. Out of eighteen observed modes, fourteen (5$\Alg$ + 9$\Eg$) are identified and assigned in agreement with the $P\bar{3}1c$ space group. Phonon energies are in a good agreement with the theoretical predictions. Two most prominent Raman modes, $\Alg^4$ and $\Alg^5$, are used to study the temperature evolution of phonon properties, and reveal three subsequent phase transitions at $T_{1}$= 142.5\;K, $T_{2}$= 190\;K and $T_{3}$= 285\;K. Furthermore, the $\Alg^5$ mode exhibits strong asymmetry, originating from enhanced spin-phonon coupling. Interestingly, the $\Alg^5$ phonon line is symmetric in the temperature range $T_{1}$--$T_{2}$, while becoming more asymmetric above $T_{3}$, indicating that the strength of spin-phonon interaction changes with temperature. We speculate, that the observed phenomenon, shown in $\Alg^5$ phonon, originates from the shift in dominance between competing magnetic states, that are found to be very close in energy \cite{PhysRevB.95.174440}.

%%%%%%%%%%%%%%%%%%%%%%%%%%%%%%%%%%%%%%%%%%
\section{Experimental and Computational Details}

The $\MnSiTe$ single crystal samples used in this study are prepared according to the procedure described in Ref. \cite{PhysRevB.98.064423}. The Raman spectra have been obtained with a Tri Vista 557 spectrometer with a 1800/1800/2400 groves/mm diffraction grating combination in a backscattering configuration. The 514 nm line of a Coherent Ar$^+$/Kr$^+$ ion laser is utilized as excitation source. The direction of the incident (scattered) light coincides with the crystallographic $c$ axis. Laser-beam focusing is achieved through a microscope objective with 50$\times$ magnification. The temperature-dependent Raman scattering measurements have been performed under high vacuum ($10^{-6}$ mbar), with the sample being placed inside of a KONTI CryoVac continuous Helium flow cryostat with a 0.5 mm thick window. The samples are cleaved in air before being placed into the cryostat. The obtained Raman spectra are corrected by a Bose factor. The spectrometer resolution is comparable to a Gaussian width of 1 \wn. 

The calculations are based on the density functional theory (DFT) formalism as implemented in Vienna Ab initio Simulation Package (VASP) \cite{PhysRevB.47.558, KRESSE199615, PhysRevB.54.11169, PhysRevB.59.1758}, with the plane wave basis truncated at a kinetic energy of 520 eV, using Perdew-Burke-Ernzehof (PBE) exchange-correlation functional \cite{Perdew1996_PRL77_3865} and Projector augmented wave (PAW) pseudopotentials \cite{Bloechl1994_PRB50_17953--17979,Kresse1999_PRB59_1758--1775}. The Monkhorst and Pack scheme of k point sampling is employed to integrate over the first Brillouin zone with $12\times12\times$10 at the $\Gamma$-centered grid. The convergence criteria for energy and force have been set to 10$^{-6}$ eV and 0.001 eV$\mathring{A}^{-1}$, respectively. The DFT-D2 method of Grimme is employed for van der Waals (vdW) corrections \cite{d2} . The vibrational modes are calculated using density functional perturbation theory implemented in VASP and Phonopy \cite{phonopy}. Previous DFT results found the energy of the ferrimagnetic state to be well above an eV per Mn below that of the non-magnetic state \cite{PhysRevB.95.174440} thus this configuration is considered in this study.

%%%%%%%%%%%%
%\begin{table}[h]
%\caption{Calculated and experimental unit cell parameters for $\MnSiTe$.}
%\label{ref:Table1}
%\begin{ruledtabular}
%\centering
%\resizebox{\linewidth}{!}{%
%\begin{tabular}{ccc}
%Mn$_3$Si$_2$Te$_6$ & Calculation & Experiment \cite{PhysRevB.98.064423}  \\ [1mm] \cline{1-3} \\[-3mm]

%$a~(\mathrm{\mathring{A}})$&& 7.046(2) \\[1mm]
%$b~(\mathrm{\mathring{A}})$& &7.046(2) \\[1mm]
%$c~(\mathrm{\mathring{A}})$ & &14.278(2)  \\[1mm]
%$\mathrm{\alpha~(deg)}$ & &90 \\[1mm]
%$\mathrm{\beta~(deg)}$  & & 90 \\[1mm]
%$\mathrm{\gamma~(deg)}$ & & 120 
%\end{tabular}}
%\end{ruledtabular}
%\end{table}
%%%%%%%%%%%%%%5

\section{Results and Discussion}

\subsection{Polarization depedence}
%%%%%%%%%%%%%%%%%%%%%%%%%%%%%%%%%%%%%%%%%%%%%%%%%%%%%%%%%%%
$\MnSiTe$ crystalizes in a trigonal ${P\bar{3}1c}$ crystal structure \cite{VINCENT1986349, PhysRevB.98.064423}. The Wyckoff positions of the atoms and their contributions to the $\Gamma$-point phonons, together with the corresponding Raman tensors, are listed in Table~\ref{ref:Table2}. In total, there are sixteen Raman-active modes (5$\Alg$ + 11$\Eg$) and seventeen infrared-active modes (6$\Asu$ + 11$\Eu$). According to the Raman tensors presented in Table~\ref{ref:Table2}, $\Eg$ symmetry modes can be observed in the Raman spectra measured in both parallel and crossed polarization configurations, whereas $\Alg$ modes arise only for those in parallel polarization configuration. 

 %WYCKOFF TABELA%%%%%%%%%%%%%%%%%%
%%%%%%%%%%%%%%%%
%%%%%%%%%%%%
\begin{table}[!htb]

\caption{Wyckoff positions of atoms and their contributions to the $\Gamma$-point phonons together with the corresponding Raman tensors for the ${P\bar{3}1c}$ space group of $\MnSiTe$. }
\label{ref:Table2}
\begin{ruledtabular}
\centering
\resizebox{\linewidth}{!}{%
\centering
\begin{tabular}{ccc}
\multicolumn{3}{c} {Space group: ${P\bar{3}1c}$} \\ \cline{1-3} \\[-2mm]

Atoms &  \multicolumn{2}{c} {Irreducible representations} \\ \cline{1-1} \cline{2-3} \\[-2mm]

Mn ($2c$) & \multicolumn{2}{c} {$A_{2g}+ A_{2u}+ E_{g}+ E_{u}$} \\[1mm]

Mn ($4f$) & \multicolumn{2}{c} {$A_{1g}+ A_{1u}+A_{2g}+ A_{2u}+ 2E_{g}+ 2E_{u}$} \\[1mm]

Si ($4e$) & \multicolumn{2}{c} {$A_{1g}+ A_{1u}+A_{2g}+ A_{2u}+ 2E_{g}+ 2E_{u}$} \\[1mm]

Te ($12i$)& \multicolumn{2}{c}{$3A_{1g} + 3A_{1u} + 3A_{2g}+ 3A_{2u}$} \\[1mm]
 &\multicolumn{2}{c}{$+ 6E_{g}+6E_{u}$}\\[1mm]
\cline{1-3}\\[-2mm]
\multicolumn{3}{c} {Raman tensors} \\ \cline{1-3} \\[-2mm]

\multicolumn{3}{c}{
$\Alg$ = $\begin{pmatrix}
a& & \\
 &a& \\
 & &b\\
\end{pmatrix}
$}
\\ [5mm]
\multicolumn{3}{c}{$
{}^1\Eg = \begin{pmatrix}
c&&\\
&-c&d\\
&d& \\ \end{pmatrix}
\;
{}^2\Eg = \begin{pmatrix}
&-c&-d\\
-c&&\\
d&& \\
\end{pmatrix}
$}
\\ [5mm]

\end{tabular}}
\end{ruledtabular}
\end{table}
%%%%%%%
 %WYCKOFF TABELA%%%%%%%%%%%%%%%%%%
%%%%%%%%%%%%%%%%
%%%%%%%%%%% 

\begin{figure}
\centering
\includegraphics[width=85mm]{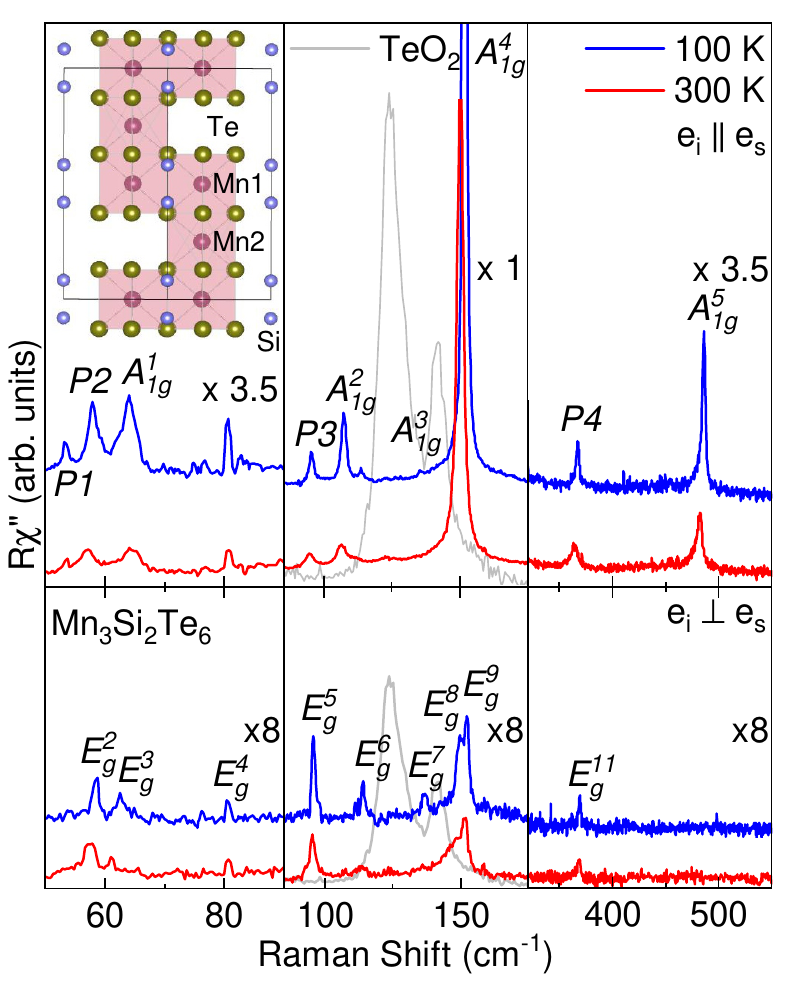}
 \caption{Raman spectra of $\MnSiTe$ single crystal measured in two scattering geometries at \textit{T}= 100 K (blue solid line) and \textit{T}= 300 K (red solid lines). Peaks observed in both geometries are identified as \Eg modes, whereas peaks observed only for the parallel polarization configuration are assigned as \Alg modes. Grey line: TeO$_2$ spectrum at 300 K, scaled for clarity. }
\label{fig:assig}
\end{figure}

As depicted in Fig.~\ref{fig:assig}, nine phonon lines are observed in parallel polarization configuration only, and identified as $\Alg$ symmetry modes. According to the symmetry analysis only five $\Alg$ symmetry modes are expected, resulting in four excess modes at 53.3 \wn, 57.9 \wn, 95.3 \wn and 366.7 \wn. These modes may arise from infra-red/silent phonons activated by disorder and from the relaxation of the symmetry selection rules \cite{PhysRevB.99.144419, Moskovits, PhysRevB.73.224401, PhysRevB.67.052405}. However, it is more likely they are overtones. Overtones are always observable in $A$ symmetries, and can become observable in Raman spectra due to disorder and/or enhanced coupling of the phonons to other excitations like in the case of spin-phonon coupling \cite{Baum2018_PRB97_054306}.  

%%%%%%%%%%%%%%%%%%%%%%%%%%%%%%%%%%%%%%%%%%%%%%
%%%%%%%%%%%%%%

Aside from the discussed $\Alg$  symmetry modes, our spectra host nine modes which obey the $\Eg$ selection rules. Therefore, nine out of the expected eleven $\Eg$ modes have been singled out and identified. The absence of two $\Eg$ modes might be attributed to their low intensity and/or the finite resolution of the spectrometer. 

Calculated and experimental phonon energies are collected in Table~\ref{ref:Table2}, and are found to be in good agreement with each other, with discrepancy being below 8\% for all observed modes.

Our data significantly differ from those presented in Ref.~\cite{doi:10.1063/5.0002168} where two Raman active modes were reported, one at 118.4 \wn and the other at 136.9 \wn, assigned as $\Eg$ and $\Alg$, respectively. The $\Eg$ and $\Alg$ modes in our spectra closest (in terms of energy) to those reported in Ref. \cite{doi:10.1063/5.0002168} are the peaks at $\sim$ 114.3 and 135.4 \wn ( Table~\ref{ref:Table2}). Although the discrepancy in phonon energy is not significant, the observed phonon linewidths strongly deviate from those presented in Ref.~\cite{doi:10.1063/5.0002168}. A possible explanation for the discrepancy is the presence of TeO$_2$ in samples presented in Ref. \cite{doi:10.1063/5.0002168}, as the peaks reported there match rather well with the Raman response of  TeO$_2$ (Fig.~\ref{fig:assig}). In order to avoid potential contamination in our study, measurements have been repeated on multiple crystals, and no oxide traces have been identified in the spectra.

\begin{table}[!htb]
\caption{Phonon symmetries and phonon frequencies of $\MnSiTe$ phonons. The experimental values are determined at 100 K. All calculations have been performed at zero 0 K. The experimental uncertainty is 0.3 \wn.} %All calculations were performed at zero temperature.}
\label{ref:Table2}
\begin{ruledtabular}
\centering
\resizebox{\linewidth}{!}{%
\centering
\begin{tabular}{ccccccc}
\multicolumn{7}{c}{Space group ${P\bar{3}1c}$} \\ \cline{1-7}  \\ [-3mm]

n$_0$ &\multicolumn{2}{c}{Symm.}  & \multicolumn{2}{c}{Exp. (\wn)} & \multicolumn{2}{c}{Calc. (\wn)} \\  \cline{1-7}   \\ [-2mm]

1&\multicolumn{2}{c}{$\Eg^{1}$} & \multicolumn{2}{c}{-} &\multicolumn{2}{c}{53.1} \\[1mm] 

2&\multicolumn{2}{c}{$P1$} & \multicolumn{2}{c}{53.3} & \multicolumn{2}{c}{-}\\[1mm] 

3&\multicolumn{2}{c}{$P2$} & \multicolumn{2}{c}{57.9}  & \multicolumn{2}{c}{-} \\[1mm] 

4&\multicolumn{2}{c}{$\Eg^{2}$} & \multicolumn{2}{c}{58.7} &  \multicolumn{2}{c}{58.5} \\[1mm] 

5&\multicolumn{2}{c}{$\Eg^{3}$} & \multicolumn{2}{c}{62.6} &  \multicolumn{2}{c}{61.8}  \\[1mm] 

6&\multicolumn{2}{c}{$\Alg^{1}$} &  \multicolumn{2}{c}{64.2}   &  \multicolumn{2}{c}{62.3}  \\[1mm]

7&\multicolumn{2}{c}{$\Eg^{4}$ }& \multicolumn{2}{c}{80.4}   &  \multicolumn{2}{c}{82.7}   \\[1mm]

8&\multicolumn{2}{c}{$P3$} & \multicolumn{2}{c}{95.3} & \multicolumn{2}{c}{-}  \\[1mm] 

9&\multicolumn{2}{c}{$\Eg^{5}$ } & \multicolumn{2}{c}{95.9} & \multicolumn{2}{c}{90.3} \\[1mm]

10&\multicolumn{2}{c}{$\Alg^{2}$} & \multicolumn{2}{c}{107.3} & \multicolumn{2}{c}{104.3} \\[1mm]

11&\multicolumn{2}{c}{$\Eg^{6}$} &  \multicolumn{2}{c}{114.0} & \multicolumn{2}{c}{106.5}  \\[1mm]

12&\multicolumn{2}{c}{$\Alg^{3}$} & \multicolumn{2}{c}{135.4} & \multicolumn{2}{c}{134.2} \\[1mm]

13&\multicolumn{2}{c}{$\Eg^{7}$} & \multicolumn{2}{c}{136.6}  & \multicolumn{2}{c}{136.1}    \\[1mm]

14&\multicolumn{2}{c}{$\Eg^{8}$} & \multicolumn{2}{c}{149.8}  & \multicolumn{2}{c}{143.4}    \\[1mm]

15&\multicolumn{2}{c}{$\Alg^{4}$} & \multicolumn{2}{c}{151.8} & \multicolumn{2}{c}{147.3}  \\[1mm]

16&\multicolumn{2}{c}{$\Eg^{9}$} & \multicolumn{2}{c}{152.6}  & \multicolumn{2}{c}{146.6}    \\[1mm]

17&\multicolumn{2}{c}{$\Eg^{10}$} & \multicolumn{2}{c}{-} &\multicolumn{2}{c}{352.7} \\[1mm] 

18&\multicolumn{2}{c}{$P4$} & \multicolumn{2}{c}{366.7}  & \multicolumn{2}{c}{-} \\[1mm] 

19&\multicolumn{2}{c}{$\Eg^{11}$} & \multicolumn{2}{c}{368.7} &\multicolumn{2}{c}{354.5} \\[1mm] 

20&\multicolumn{2}{c}{$\Alg^{5}$} & \multicolumn{2}{c}{486.7}  &\multicolumn{2}{c}{475.83}  \\

\end{tabular}}
\end{ruledtabular}
\end{table}

\subsection{Temperature dependence}

Some of the modes represented in Fig.~\ref{fig:assig} exhibit an asymmetric lineshape. Although, the appearance of a mode asymmetry can be attributed to the presence of defects \cite{Lazarevi__2013} this would have a significant impact also on the line widths of other modes in spectrum, which is not the case here. The asymmetry may arise from phonon-continuum coupling, \textit{e.g.} spin-phonon. The lineshape originating from such a coupling is given by the Fano profile \cite{PhysRev.124.1866}:

\begin{center}
$\mathrm{I(\omega)=I_0\frac{(q+\epsilon)^2}{1+\epsilon^2}}$,
\end{center}
where $\epsilon$($\omega$)=2($\omega$-$\omega_0$)/$\Gamma$. Here, $\omega_0$ is the phonon frequency in the absence of interaction, $\Gamma$ is the full width at the half maximum (FWHM), $I_0$ is the amplitude and q is the Fano parameter. The Fano parameter and FWHM depend on the interaction strength between the phonon and the continuum, and therefore can be used as its indicator. To include the finite spectral resolution of the experimental setup, the Fano profile is convoluted with a Gaussian function as demonstrated in Ref. \cite{Baum2018_PRB97_054306}. 

%%%%%%%%%%%%%%%%%%%%%%%%%%%%%%%%%%%%%%%%%%%%%%%%%%%%%%%%%%

\begin{figure}[h!]
 \centering
  \includegraphics[width=85mm]{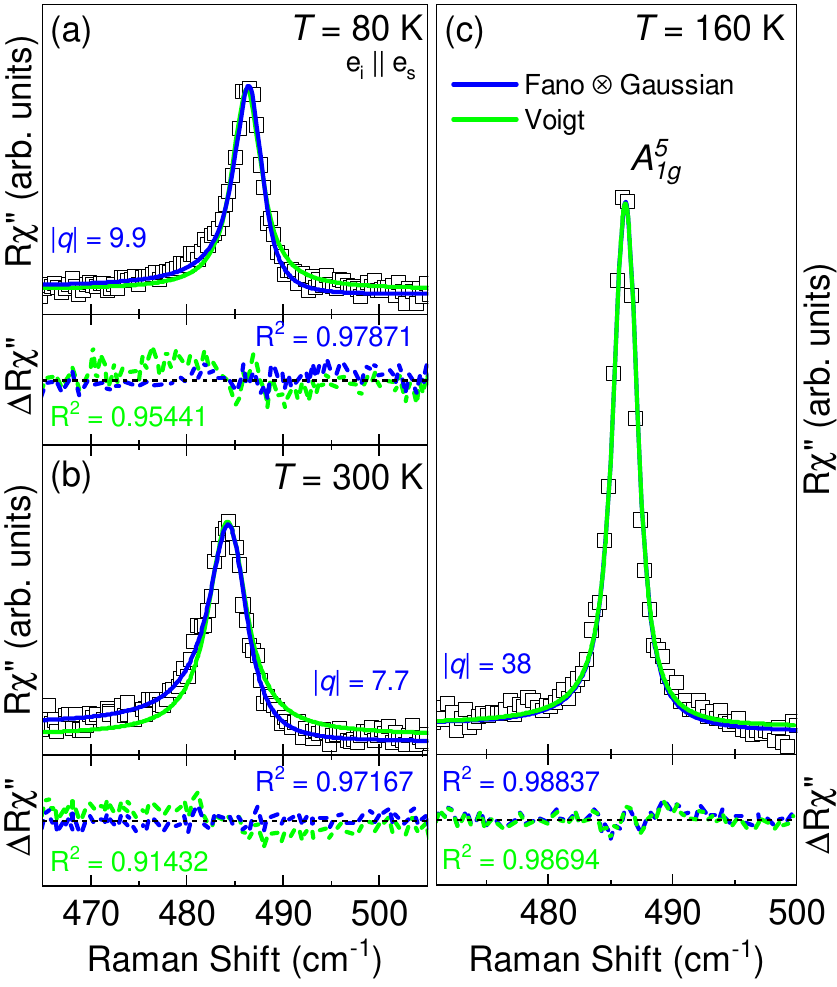}
  \caption{Raman response as a function of the Raman shift. Quantitative analysis of the $\Alg^5$ mode at temperatures as indicated. (a) and (b) The blue solid lines represent the line shape obtained as a convolution of Fano profiles and Gaussian, whereas the green solid lines represent Voigt profiles. (c) Comparison between asymmetric (deep blue) and symmetric (light blue) line shapes obtained as a Fano-Gaussian convolution and a Voight profile. Experimental data is represented by open squares.}
 \label{fig:comp}
\end{figure}
%%%%%%%%%%%%%%

The high intensity peak at 486.7 \wn, identified as the $\Alg^{5}$ symmetry mode, does not overlap with any other mode. The quantitative analysis of this peak is performed using both the symmetric Voigt profile and the Fano-Gaussian convolution mentioned above.  The comparison between two models and the experimental data at 80\;K and 300\;K are presented in Fig.~\ref{fig:comp} (a) and (b), respectively. The asymmetric lineshapes provide a satisfactory description of the measured phonon line shape, suggesting the presence of an additional scattering mechanism in \MnSiTe. 
 
\begin{figure}
 \centering
  \includegraphics[width=85mm]{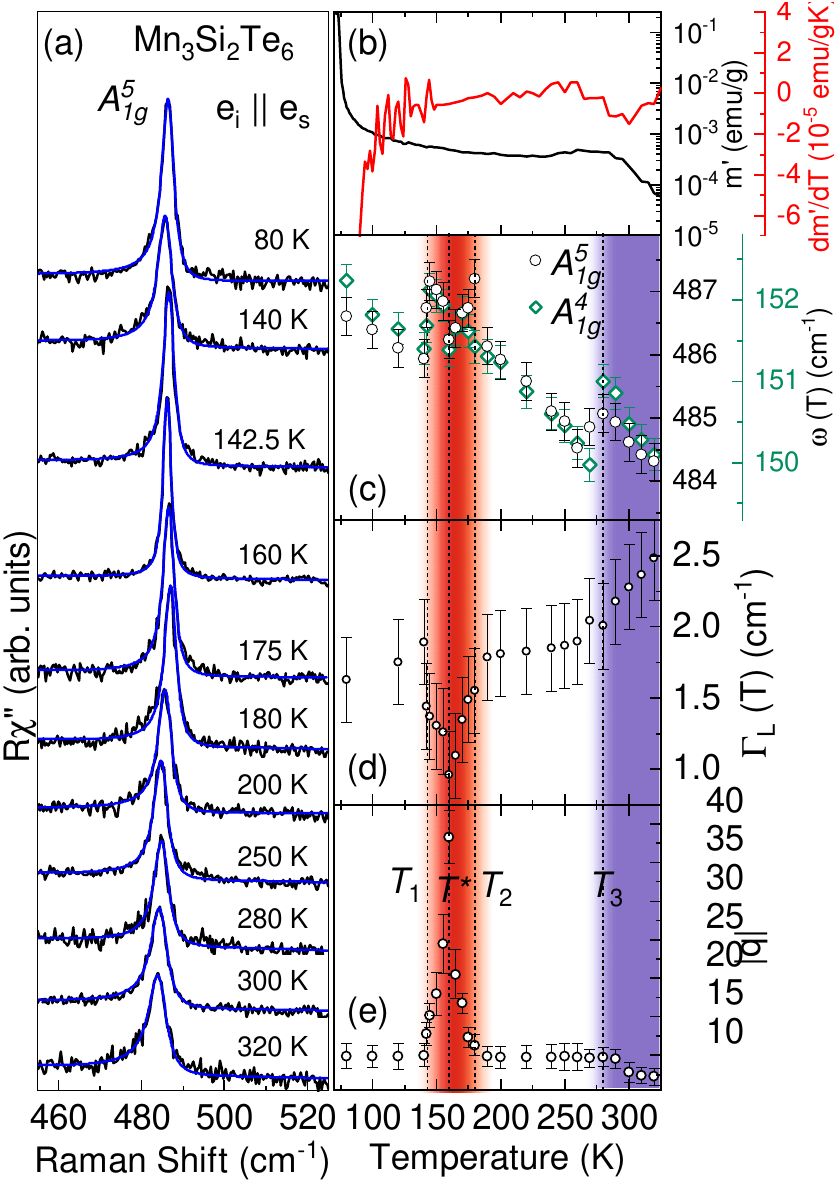}
  \caption{(a) The spectral region of the $\Alg^5$ Raman-active mode of $\MnSiTe$ at indicated temperatures measured in the parallel polarization configuration. Green solid lines represent line shapes obtained as a convolution of the Fano line shape and Gaussian, calculated to fit the experimental data. (b) Temperature dependence of ac susceptibility real part \textit{m'(T)} and its temperature derivative ploted as a function of temperature with $\mathbf{H}$ $||$ $\mathbf{ab}$. Temperature dependence of (c) the energy of the $\Alg^4$ and $\Alg^5$ as well as (d) the line width, and (e) the Fano parameter $|q|$ of the $\Alg^5$ mode. }
 \label{fig:a8g}
\end{figure}

The spectral region of the $\Alg^5$ Raman-active mode in the temperature range of interest is presented in Fig.~\ref{fig:a8g} (a). The blue solid lines represent fits to the experimental data obtained using the Fano-Gaussian line shape. Temperature dependence of the phonon energy, line width, and the Fano parameter $|q|$ of the $\Alg^5$ mode are depicted in Fig.~\ref{fig:a8g} (c), (d) and (e), respectively. By increasing the temperature above 80\;K, the $\Alg^5$ mode broadens and softens up to $T_{1}$= 142.5\,K, where it abruptly narrows and shifts to higher energies followed by further softening and narrowing up to \Tst= 160\;K. Additional heating leads to a broadening and hardening before the drop in phonon energy at $\sim$ $T_{2}$= 190\;K. In the region $T_{2}$ the mode softens and broadens with additional jump in phonon energy at $T_{3}$= 285\;K. A similar trend is also observed for the $\Alg^4$ mode, as evidenced in Fig.~\ref{fig:a8g} (b).  
 
%%%%%%%%%%%%%%5
%%%%%%%%%%%%%%

This intriguing temperature dependence is also manifested in the asymmetry \textit{i.e.} Fano parameter $|q|$ [Fig.~\ref{fig:a8g} (d)] of the $\Alg^5$ peak. At the lowest experimental temperature, 80\;K , the $\Alg^5$ mode exhibits strong asymmetry with a Fano parameter $|q|$= 9.9. Upon heating the sample to $\sim$ $T_{1}$ a Fano parameter remains nearly constant before the significant increase in the temperature range between $T_{1}$ and \Tst resulting in a symmetric lineshape ($|q|$= 38, Fig.\ref{fig:a8g} (c)). Further temperature increase leads to a strong decrease of $|q|$ up to $T_{2}$, where the asymmetry is restored ($|q|$= 9.9) remaining almost constant up to $T_{3}$. At higher temperatures, the lineshape becomes more asymmetric, reaching $|q|\sim$8 at the highest experimentally accessible temperature $T$= 320\,K.

While the ferrimagnetic order in $\MnSiTe$ is established only at $T_c$= 78\;K \cite{PhysRevB.98.064423, doi:10.1063/5.0002168}, the asymmetry of the mode can be traced to enhanced spin-phonon interaction related to short-range correlation \cite{doi:10.1021/acs.inorgchem.0c02060, Sandilands_2015, PhysRevB.99.144419}, that can survive up to temperatures well above $T_c$, similarly to $\MnSiTe$'s 2D analog $\CST$ \cite{Ana}. Based on the results presented in Ref.~\cite{PhysRevB.95.174440}, we may speculate that these short-range correlations are likely in terms of the antiferomagnetic dimers and/or trimers between different Mn layers, Mn1 and Mn2 (as depicted in Fig. \ref{fig:assig}) in the paramagnetic background. However, this alone cannot explain sudden changes in the properties of the $\Alg^5$ phonon mode. Rather the existence of competing short-range magnetic phases may be responsible for the observed behavior of the phonon modes. The first phonon mode anomaly at $T_{3}$= 285\;K corresponds to the anomaly in $m'(T)_{ab}$ [Fig.~\ref{fig:a8g} (b)] and can be seen as the outlet of additional short-range order in the paramagnetic domains \cite{PhysRevB.103.245122} and possibly change of their nature of previously established ones. The onset in temperature with the magnetization anomaly near 330\;K \cite{ doi:10.1063/5.0002168, Ni_2021} is likely the consequence of local disorder.
At $T_2$ $\MnSiTe$ becomes locally magnetically frustrated resulting in the change in magnetostriction and rapid decrease of spin-phonon interaction that is manifested in the strong evolution of the phonon self-energy (Fig.~\ref{fig:a8g}). At this temperature both magnetoresistance and nonlinearety of Hall resistance become observable \cite{Ni_2021}.    In this scenario, by further lowering the temperature, at $T_1$ new short range magnetic order, likely antiferromagnetic, is established and the strong spin-phonon interaction is established. In order to fully understand the complex evolution of the short-range magnetic correlation in $\MnSiTe$ that is manifested through the anomalous temperature development of  $\Alg^5$ mode, further investigations are required.

\section{Conclusion}
\label{sec:conclusion}

The lattice dynamic in single crystalline $\MnSiTe$ using Raman spectroscopy in analyzed. Five $\Alg$ modes and nine $\Eg$ modes are observed and assigned according to the ${P\bar{3}1c}$ symmetry group. Four additional peaks to ones assigned to ${P\bar{3}1c}$ symmetry group, obeying $\Alg$ selection rules, are attributed to overtones. The pronounced asymmetry of the $\Alg^5$ phonon mode at 100\,K and 300\,K. The unconventional temperature evolution of the $\Alg^5$ Raman mode reveal three successive, possibly magnetic, phase transitions that significantly impact strength of the spin-phonon interaction in $\MnSiTe$.  These are likely caused by the competition between the various magnetic states, close in energy.
This study provides a comprehensive insight to the lattice properties, their temperature dependence and shows arguments for existence of the competing short-range magnetic phases in \MnSiTe.

\section*{Acknowledgements}
The authors acknowledge funding provided by the Institute of Physics Belgrade, through a grant from the Ministry of Education, Science and Technological Development of the Republic of Serbia, Project F-134 of the Serbian Academy of Sciences and Arts, the Science Fund of the Republic of Serbia, PROMIS, 6062656, StrainedFeSC, Austrian Science Fund (FWF) through the Project No. P31423, and the support of Austrian Academy Of Sciences’ Joint Excellence In Science And Humanities (JESH) Program (JP). DFT calculations were performed using computational resources at Johannes Kepler University (Linz, Austria). Materials synthesis was supported by the U.S. DOE-BES, Division of Materials Science and Engineering, under Contract DE-SC0012704 (BNL).  

\bibliography{reference}

%\end{document}

%%%%%%%%%%%%%%%%%%%%%%%%%%%%%%%%%%%%%%%%%%%%%%%%

%\begin{appendix}
%\label{sec:appendix}

%\setcounter{figure}{0}
%\renewcommand\thefigure{\thesection.\arabic{figure}}
%\renewcommand\thefigure{A\arabic{figure}}

%\setcounter{table}{0}
%\renewcommand\thetable{A\Roman{table}}

%\section{IR-active phonons}

%IR active phonon phonon energies numerically calculated for ${P\bar{3}1c}$ $\mathrm{VI_3}$ (half-metal) are summarized in Table~\ref{ref:TableA1}.
%\label{Asec:IRphonons}
%%%%%%%%%

%\begin{table}[h]
%\caption{Numerically calculated (at zero temperature) phonon energies in for the ${P\bar{3}1c}$ (half-metal) structure of $\mathrm{VI_3}$ for IR active phonons.}
%\label{ref:TableA1}
%\begin{ruledtabular}
%\centering
%\begin{tabular}{cc}
%\multicolumn{2}{c} {Space group  $P\bar{3}1c$ (half-metal)} \\ [1mm] \cline{1-2}  \\[-1mm]
 %Symm. & Calc. (\wn) \\ [1mm] \cline{1-1} \cline{2-2} \\[-0.5em]
%$\Eu^{1}$ & -5.8 \\[1mm] 
%$\Asu^{1}$ & 21.9  \\[1mm]
%$\Eu^{2}$ & 60.2 \\[1mm]
%$\Eu^{3}$ & 72.1 \\[1mm]
%$\Asu^{2}$  & 74.1 \\[1mm]
%$\Eu^{4}$ & 88.5 \\[1mm]
%$\Asu^{3}$ & 92.3  \\[1mm]
%$\Eu^{5}$& 95.7\\[1mm]
%$\Eu^{6}$ & 104.3  \\[1mm]
%$\Asu^{4}$ & 148.2  \\[1mm]
%$\Eu^{7}$  & 158.7  \\[1mm]
%$\Eu^{8}$ & 162.2 \\[1mm]
%$\Asu^{5}$ & 184.1 \\[1mm]
%\end{tabular}
%\end{ruledtabular}
%\end{table}
%\vfill
%\end{appendix}

\end{document}